# Tornado lift
## Alexander Ivanchin

*It is shown that one of the causes for tornado is Tornado Lift. At increasing vortex diameter its kinetic energy decreases to keep the moment of momentum constant. A kinetic energy gradient of such vortex is Tornado Lift. Evaluation shows that contribution of Tornado Lift in air lifting in a tornado is comparable to buoyancy according to the order of magnitude.*

At the present there is no satisfactory tornado theory. The authors of present hypothesis endeavour to give a qualitative description without consecutive quantitative evaluations in spite of the fact that the laws of gas mechanics are well known. They have not taken into consideration on important fact, that is to say, Tornado Lift (TL).

As moments of forces of mechanics, electric, magnetic or other nature have not been observed which rotate the air of tornado, there is nothing left to do but explain this phenomenon applying the law of conservation of moment of momentum. Owing to the earth revolution atmosphere has moment of momentum.

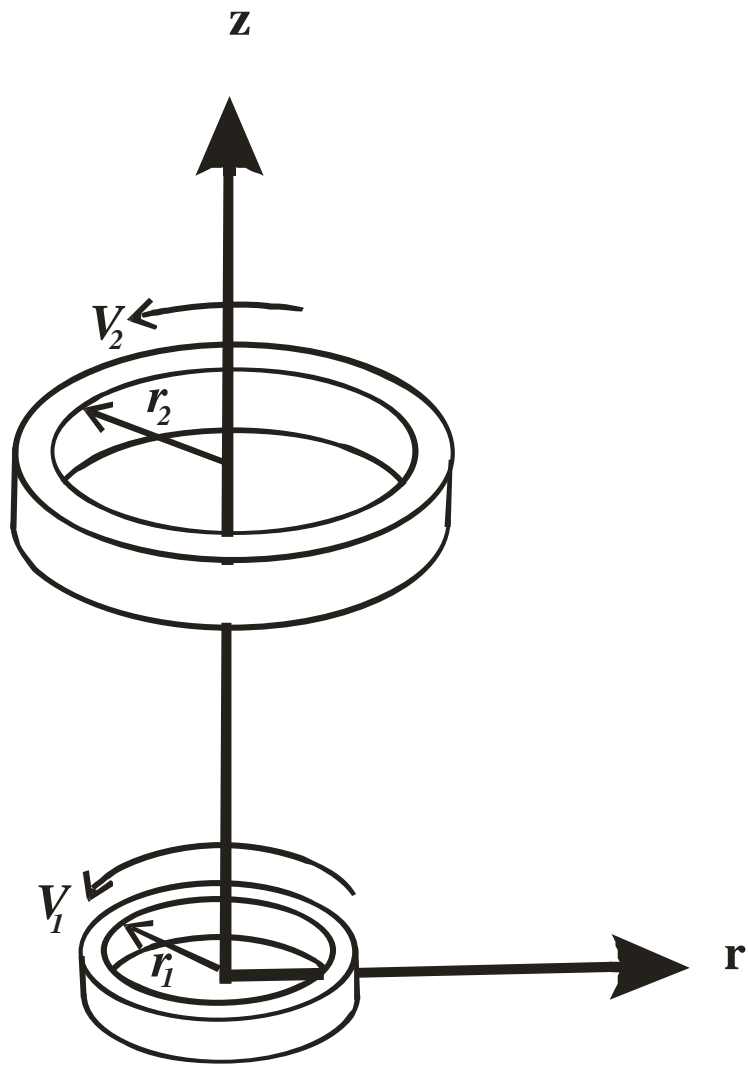

Figure 1. It is diagram for a gas ring with a rotation velocities $V_1$, $V_2$ $(V_1 > V_2)$ and radii $r_1$, $r_2$ $(r_1 < r_2)$ at lifting to certain altitude. The ring radius increases on account of height.

Rotating gas ring (figure 1) has the following moment of momentum (per mass unit)

$$M = Vr \qquad (1)$$

Here, $V$, $r$ are an azimuthal velocity and a ring radius, respectively. In the case of absence of force moments, $M$ is the preserving value i.e. $M = \text{const}$. According to (1) at convergent flow velocity can be arbitrarily large. This mechanism was investigated by Kuo [1].

In order to raise rotating air we should apply certain force. The nature of this force is following. The rotating gas ring has kinetic energy (per mass unit):

$$W = \frac{V^2}{2} = \frac{M^2}{2r^2}$$

If the ring radius depends on the height $z$, i.e. $r = r(z)$, kinetic energy depends on the height as well. According to physics laws, a generalized coordinate derivative from the energy is the generalized force:

$$f = \frac{dW}{dz} = -\frac{M^2}{r^3}\frac{dr}{dz} \qquad (2)$$

For tornado, as observations show, the radius of vortex increases with height due to decreasing air pressure and density and other causes, i.e. $dr/dz > 0$, and consequently kinetic energy of vortex decreases with height. A force has direction towards a decreasing side of energy, i.e. upwards in this case. Let call this force Tornado Lift (TL).

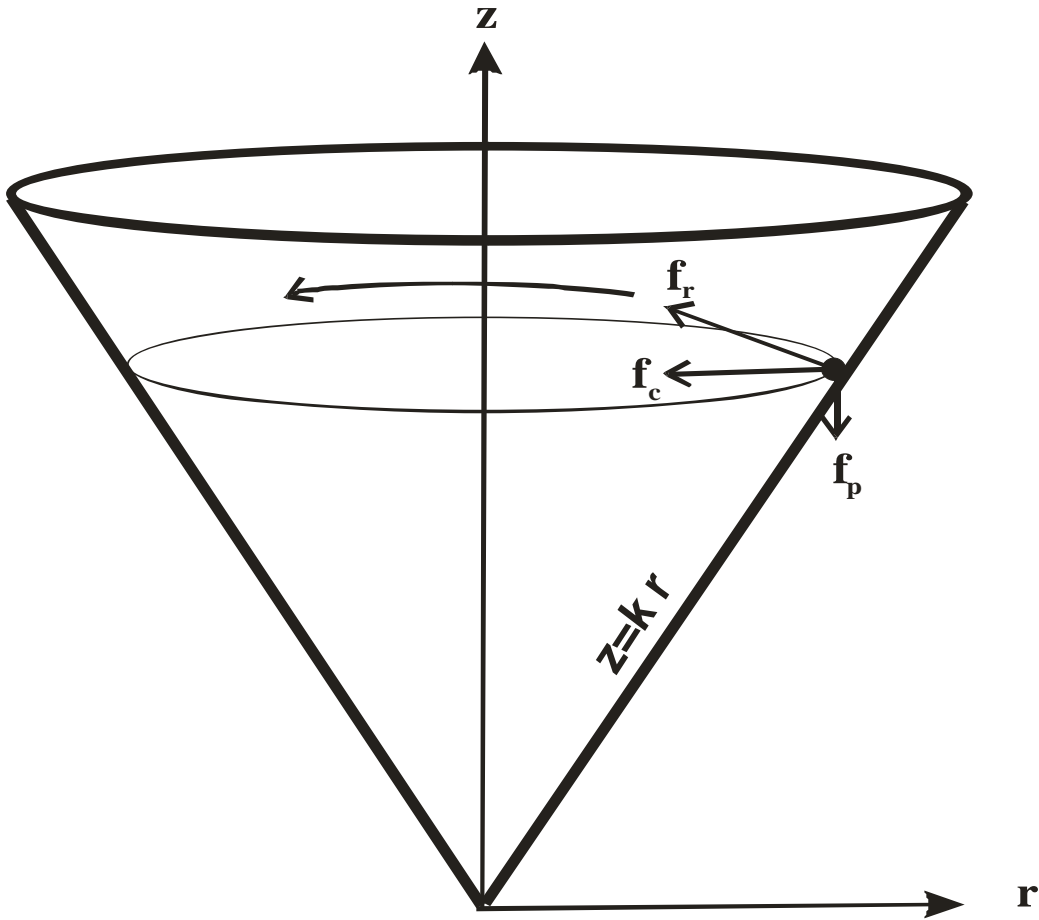

Figure 2. It is diagram for a material point moving on the surface of cone.

In order to show TL well, let us consider such problem: on the surface of cone $z = kr$ with no friction a mater point of a unit mass moves automatically (Figure 2). As there are no any moments of external force here, the moment of momentum $M = V z/k$ is constant, and kinetic energy is equal to $W = k^2 M^2 / 2z^2$. Here, as for a gas ring at increasing height the radius of rotation becomes larger that decreases kinetic energy. The Hamiltonian function is written in the following form $H = W + gz$, and the force acting in the point upwards is as follows: $F = dH/dz = dW/dz + g$. Here $g$ is free fall acceleration. The summand $dW/dz = -k^2 M^2 / z^3$ is an analogue for TL. The same result will be in a force approach if we consider the sum of forces as centripetal $\mathbf{f}_c$, the weight $\mathbf{f}_p$, and reaction at the support $\mathbf{f}_c$.

In the present mechanics there are two approaches: a force one and energetic. At a force approach it is necessary to show the forces acting between bodies; an energetic approach is based on observing the energy of the system, and forces are identified as generalized coordinate derivatives from energy. Historically a force approach is the first one, but an energetic approach is more universal. Thermodynamics, statistical physics and quantum mechanics are based on the latter. These two approaches coincide with each other if the system consists of discrete bodies. However, in case of continuous medium, it is not always possible to consider the problem in a force approach. Typical representatives of such forces are surface tension force (an interphase surface area derivatives from energy) and osmotic pressure (a concentration derivative from energy) and so on. Among this kind of forces there is TL.

As my experience in discussing this work shows, it is this peculiarity of TL that can not be understood that is connected with an almost exclusive application of a force approach in atmosphere dynamics. If we keep strictly to the fact, that every problem should be solved exclusively in a force approach, we shall repudiate the bulk of thermodynamics, quantum and statistic mechanics. I think that the nature of a number of atmospheric phenomena can be understood via a close consideration of power, that allows to take into account the forces which can not be studied using a force approach.

Let us consider energy of a gas particle in atmosphere. Its total energy per mass unit is as follows:

$$E = H + gz + P + W \tag{3}$$

Here $H$ - an enthalpy of a gas particle, $W$ - its kinetic energy, $P$ - pressure function for the particle environment. Then we differentiate (3) with respect to the height, find the force acting on the particle

$$\frac{dE}{dz} = \frac{dH}{dz} + \left(g + \frac{1}{\rho}\frac{dp}{dz}\right) + \frac{dW}{dz} \tag{4}$$

Here, $p$ - atmospheric pressure, $\rho$ - air density. The sum in the parenthesis is buoyancy force, the last summand is TL. Let us evaluate these forces using the following example. In order to raise a gas particle at an altitude of 1 km above sea level in standard atmosphere by means of buoyancy, it should be heated about $\Delta T \sim 4\ K$, i.e. add to it some enthalpy $C_p \Delta T \sim 4000\ J/kg$. If an azimuthal velocity for the vortex close to ground surface $\sim 100\ m/s$ and at height of 1 km the vortex radius increases three times, its kinetic energy decreases by $4500\ J/kg$. Hence, if rotation velocity is large, gas particle can be raised at an altitude of about kilometers using kinetic energy of rotation. In paper [2] the authors show the connection of tornado genesis and kinetic energy of mesocyclone. If in a mesocyclone convection presents, that automatically leads to convergent flow and thereby according to (1) to the increase of an azimuthal velocity of rotation and TL as well. TL gives increasing contribution into the air raising, increasing the velocity of lifting, and thereby convergences. The process becomes auto accelerating until it turns into a tornado. That is to say, convective flows serve for tornado starting. One of the interesting ideas is evaluation of critical conditions satisfying which the formation of the tornado is possible. Frequently tornado is proceeded by tornado cloud. According to present suggestions appearance of such a cloud in connected with convective lift of wet air. As rising air rotates the tornado cloud rotates as well that can be seen. A tornado cloud can not generate tornado, as they speak sometimes, it in its turn, is a formation of convective processes with rotation. Condensation of moisture makes them visible.

Theoretically a tornado was studied by Staley et al. [3,4]. In these works different aspects of vortex flow have been studied but in the initial set of equations TL has not been taken into consideration, which should be put into, in particular, the equation for movement together with viscosity, pressure and gravity force. Above mentioned that influence of TL for real flows is comparable to buoyancy, therefore TL should be applied in a corrective theory.

A suggestion is widely known that a tornado is formed by turning of a horizontal vortex into a vertical one. A vortex with a horizontal axis is formed on account of shift of horizontal component for wind velocity. This hypothesis is corroborated by a number of authors. However, this hypothesis has questions which fall short of physics, namely

1. Where do moments rotating air originate from? According to the low of conservation of momentum angular acceleration is proportional to the moment applied. What kind of nature is for this moment? It is necessary to take into consideration the fact that on account of the pressure differential a moment of force can not be formed at all as pressure is a potential function. If moment does not present, rotation does not occur. A vertical shift of a horizontal component for wind velocity is not rotation. Formation of vortex presents on account of viscosity forces at interaction with streamline surface, but rotation velocity can not exceed flow velocity, as friction is a dissipative process by its nature. Therefore the cause for a high air rotation during rotation around a horizontal axis has not been explained in this hypothesis.

2. Owing to what moment rotation axis is turned from horizontal into vertical? For this is necessary to apply moment perpendicular to rotating axis. Convective flows, even if they form a moment, can not turn in principle a horizontal rotational axis into a vertical one, they can only change axis orientation in a horizontal plane according to the rule of moments in mechanics.

There are other contradictions in the hypothesis of a horizontal vortex.

The mechanics of TL acting offered in this work does not pretend to the entire quantitave description for such a complex phenomenon a tornado. Here the author proves the necessity of taking into account TL.